\documentclass{PoS}
\def\lesssim{\buildrel < \over {_{\sim}}}
\def\gtrsim{\buildrel > \over {_{\sim}}}
\def\HH{H$_2$\ }
\def\sigmav{\langle\sigma v\rangle}

\title{WIMP annihilation effects on primordial star formation}

\ShortTitle{DM annihilation in PopIII formation: chemical feedback}

\author{E. Ripamonti\thanks{Universit\`a degli Studi dell'Insubria,
    Dipartimento di Fisica e Matematica, Como, Italy},
F. Iocco\thanks{INAF/Osservatorio Astrofisico di Arcetri, Firenze,
    Italy},
A. Bressan\thanks{INAF/Osservatorio Astronomico di Padova, Padova,
  Italy; INAOE, Puebla, Mexico; SISSA, Trieste, Italy},
R. Schneider\thanks{INAF/Osservatorio Astrofisico di Arcetri, Firenze,
    Italy},
A. Ferrara\thanks{SISSA, Trieste, Italy},
P. Marigo\thanks{INAF/Osservatorio Astronomico di Padova, Padova, Italy}
 \\}

\abstract{We study the effects of WIMP dark matter (DM) annihilations on
  the thermal and chemical evolution of the gaseous clouds where the
  first generation of stars in the Universe is formed.

  We follow the
  collapse of the gas inside a typical halo virializing at very high
  redshift, from well before virialization until a stage where the
  heating from DM annihilations exceeds the gas cooling rate.  The DM
  energy input is estimated by inserting the energy released by DM
  annihilations (as predicted by an adiabatic contraction of the
  original DM profile) in a spherically symmetric radiative transfer
  scheme. In addition to the heating effects of the energy absorbed, we
  include its feedback upon the chemical properties of the gas, which is
  critical
  to determine the cooling rate in the halo, and hence the
  fragmentation scale and Jeans mass of the first stars.

  We find that DM annihilation {\it does} alter the free electron and
  especially the \HH fraction when the gas density is
  $n\gtrsim$10$^4\#$/cm$^3$, for our fiducial parameter values. However,
  even if the change in the \HH abundance and the cooling efficiency of
  the gas is large (sometimes exceeding a factor 100), the effects on
  the temperature of the collapsing gas are far smaller (a reduction by
  a factor $\lesssim 1.5$), since the gas cooling rate depends very
  strongly on temperature: then, the fragmentation mass scale is reduced only
  slightly, hinting towards no dramatic change in the initial mass
  function of the first stars.
}

\FullConference{Identification of dark matter 2008\\
		 August 18-22, 2008\\
		 Stockholm, Sweden}

\begin{document}

\section{Introduction}
In the currently favoured $\Lambda$CDM cosmological model (see
e.g. Komatsu et al. 2009), the bulk of the matter component is believed
to consist of unknown particles, commonly called Dark Matter
(DM). Weakly Interacting Massive Particles (WIMPs) are among the
favorite candidates complying with constraints from cosmology and
particle physics (see e.g. Bertone et al. 2005). As the lightest
supersymmetric partners of the standard model in theories with R--parity
conservation, WIMPs are stable Majorana fermions {\it weakly} coupled to
baryons, and are their own antiparticles.

DM drives the process of structure formation, dominating the large-scale
gravitational potential of galaxies; but DM density is too small for the
energy released by annihilations to affect the evolution of the Inter
Galactic Medium, or stellar formation and evolution in the present-day
Universe (the central parsecs of the Milky Way are a possible exception;
see Scott et al. 2009).

The formation of the first generation of stars in the Universe is
believed to be very different from all subsequent star formation
episodes: the unique conditions of the gas in the star-forming halos at
redshift z$\gtrsim$20 (above all, the absence of metals, which implies
that gas cooling depends on \HH molecules) seem to indicate that
typically only one massive star forms at the very center of each halo
(see e.g. the proceedings of the ``First Stars III'' conference).
However, Spolyar, Freese and Gondolo (2008, hereafter SFG) suggested
that these unique conditions may enhance the effects of DM
annihilations, altering primordial star formation.

SFG showed that the formation of a protostellar cloud at the center of
an halo (predicted by the standard scenario) would largely increase the
central DM density through the adiabatic contraction mechanism (AC;
Blumenthal et al. 1986).  This would boost the DM annihilation rate, and
the energy deposited in the surrounding gas would equal the one radiated
away by \HH cooling by the time gas density exceeds
$n\equiv\rho/m_p\gtrsim10^6-10^{10}\,{\rm cm^{-3}}$ (depending on parameters).  They
dubbed a {\it dark} star the object resulting from an equilbrium between
DM annihilation and gas cooling. Natarajan et al. (2009) confirmed these
results by using DM and gas profiles from three-dimensional cosmological
simulations of first star formation, relying on extrapolations of the
inner DM profile. Freese et al. (2008ab; see also the contributions of
Spolyar and Gondolo in these proceedings) and Iocco et al. (2008)
studied the evolution of an hydrostatic object powered by DM
annihilation, starting from an initial central density
$n\sim10^{16}\,{\rm cm^{-3}}$. Their initial conditions and techniques
differ, and we address the reader to the original papers for further
details.

In these proceedings we study the yet unexplored phases between
virialization and the formation of an hydrostatic object,
investigating whether the energy released by DM annihilation
has an effect on the cooling of the gas, and on the Jeans mass and final
mass range of primordial stars.

%#####

\section{Method and results}\label{method}

%#####

We follow the gas evolution introducing a feedback between gas
contraction, DM annihilation, energy absorption and gas chemistry; we do
so by means of a lagrangian 1D spherically symmetric
code including the treatment of gravitation, hydrodynamics, chemistry
and cooling (see Ripamonti et al. 2007).
We have modified it by introducing an AC algorithm
(from Gnedin et al. 2004) accounting for the modification of
the DM profile due to the extra pull from baryons, as they collapse to
the center of the halo.
The energy per unit time per unit volume released by the DM annihilation
reads:
\begin{equation}
l_{dm}(r)=f c^2\frac{\sigmav}{m_{DM}}\rho_{DM}^2(r),
\label{DMlum}
\end{equation}
where $\rho_{DM}$ is the local DM density, and we adopr
$m_{DM}$=100GeV/c$^2$, $\sigmav$=3$\times$10$^{-26}$cm$^3$/s for the
neutralino mass and its self-annihilation rate.  The factor $f$=2/3
accounts for the fraction of energy carried away by neutrinos (typically
1/3) at the end of the shower induced by annihilation primaries (see
Bertone et al. 2005). However, this energy is in the form of photons and
stable charged particles (electrons or positrons) with hard spectra and
cutoff at the neutralino mass. We estimate the energy actually absorbed
by the baryons through the code radiative transfer algorithm, assuming
an effective frequency integrated opacity $\kappa$=0.01 cm$^2$/g. This
value (which might be treated as a parameter) does not come from a
rigorous treatment; however, it is in rough agreement with the treatment
from SFG, thus allowing for result comparison.

\begin{figure}
\centering
\includegraphics[angle=0,width=0.48\textwidth]{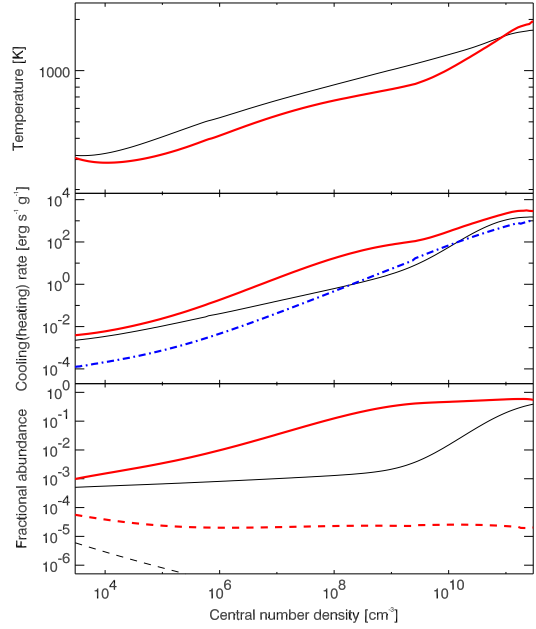}
\caption{Evolution of the central shell of the simulated object, as a
  function of its density. Top panel: temperature (solid); middle
  panel: gas cooling rate (solid) and heating rate from DM annihilations
  (dot-dashed);
  bottom panel: abundances of \HH (solid) and $e^-$ (dashed). Thick
  (red) lines refer to the case with DM annihilation, thin (black) ones
  to the ``standard'' case with no DM annihilations.}
\label{fig:fig1}
\end{figure}

We consider a typical halo with mass 10$^6$M$_\odot$, with a gas
fraction $0.175$, virializing at $z_{vir}=20$. We start integrating
at $z=1000$, from flat DM and gas profiles; the code follows the
gas evolution, whereas the DM evolution is taken from a smooth recipe
(based on analytical theories of structure formation) which slowly
transforms a flat DM profile into an appropriate ($R_{200}\simeq R_{\rm
  vir}\simeq5\times10^{20}\,{\rm cm}$, $c=10$) NFW profile (Navarro et
al. 1996) at $z=z_{vir}$; after virialization, the DM profile remains
fixed, with the exception of the central regions, where its shape is
modified by AC.

In Fig. \ref{fig:fig1} we compare the evolution of the central
properties of models with (thick lines) and without (thin lines) DM
annihilation effects, as a function of the gas central density.  In most
of the range we show, the DM energy input induce a {\it reduction} of
the gas temperature (top panel).  This is a consequence of the enhanced
cooling rate (central panel, solid), induced by the increase in the \HH
abundance (bottom panel, solid), which in turn is due to the catalyzing
effects of the increased free electrons abundance (bottom panel, dashed)
on the the \HH formation chain (H + $e^-\rightarrow$ H$^-$ + $\gamma$;
H$^-$ + H$\rightarrow$ H$_2$ + $e^-$). However, the temperature drops
only by a factor $\lesssim 1.5$, much smaller than the increase in the \HH
abundance: in fact, given
the strong dependence of \HH cooling from temperature, a modest
temperature reduction can balance a much larger increase in the \HH
abundance.

In the model with annhilations, when $n\gtrsim 10^9\rm{cm^{-3}}$ most of
H is already molecular, and \HH cannot increase any more: the cooling
rates of the two models slowly converge, as H is converted into \HH also
in the standard case.  Furthermore, as the central gas and DM densities
increase, direct heating from annihilations (central panel, dot-dashed)
becomes significant, and finally takes central temperature over that of
the standard case ($n\gtrsim 10^{11}\rm{cm^{-3}}$).

%#####
\section{Conclusions}
%#####
We present results from the first calculation of the primordial
star formation process which self-consistently includes gas
hydrodynamics, chemistry, and DM annihilation.
We show that the feedback between DM annihilations and chemistry
{\it does} change the gas thermodynamics. However, this appear
to have limited effects on the fragmentation scale of primordial
clouds, which should be reduced by a factor $\lesssim 2$ (if the fragment
mass is $\propto M_{Jeans} \propto T^{3/2}$).
Although further study is definitely necessary (see the forthcoming
Ripamonti et al. 2009), our results strongly
hint that DM annihilation effects do not alter dramatically
the fragmentation during the first star formation episode in the Universe.
A. B. and P. M. acknowledge contract ASI I/016/07/0,
F.I. is supported by MIUR through PRIN-2006.

\end{document}